\documentclass[useAMS,usenatbib]{mn2e}

\usepackage{epsfig,amsmath,amssymb,amsfonts,mathrsfs,latexsym,graphicx,lscape}
\usepackage{times}
\usepackage{longtable}
\usepackage{xcolor}
\input{epsf}
\usepackage{multirow}
\usepackage{makecell}
\usepackage{natbib}
\usepackage{color}
\usepackage{subfigure}
\usepackage{url}

\title[Disc--corona interaction]
{Disc--corona interaction in the heartbeat state of GRS 1915+105}

\author[Yan et al.]
       {Shu-Ping Yan$^{1,2,3}$\thanks{E-mail: yanshuping@pmo.ac.cn}, 
        Li Ji$^{1,2}$, 
        Si-Ming Liu$^{1,2}$, 
        Mariano M\'endez$^{4}$, 
        Na Wang$^{5}$, 
\newauthor 
        Xiang-Dong Li$^{3,6}$, 
        Jin-Lu Qu$^{7}$, 
        Wei Sun$^{1,2}$, 
        Ming-Yu Ge$^{7}$, 
        Jin-Yuan Liao$^{7}$,
\newauthor 
        Shu Niu$^{1,2}$, 
        Guo-Qiang Ding$^{5}$, 
        Qing-Zhong Liu$^{1}$\\ 
      $^{1}$Purple Mountain Observatory, Chinese Academy of Sciences, Nanjing 210008, China\\
      $^{2}$Key Laboratory of Dark Matter and Space Astronomy, Chinese Academy of Sciences, Nanjing 210008, China\\
      $^{3}$Key Laboratory of Modern Astronomy and Astrophysics (Nanjing University), Ministry of Education, Nanjing 210093, China\\
      $^{4}$Kapteyn Astronomical Institute, University of Groningen, P.O. Box 800, 9700 AV Groningen, The Netherlands\\
      $^{5}$Xinjiang Astronomical Observatory, Chinese Academy of Sciences, Xinjiang 830011, China\\
      $^{6}$Department of Astronomy, Nanjing University, Nanjing 210093, China\\
      $^{7}$Key Laboratory for Particle Astrophysics, Institute of High Energy Physics, Chinese Academy of Sciences, Beijing 100049, China}

\begin{document}

\date{Accepted 2017 November 3. Received 2017 October 19; in original form 2017 March 24}

\pagerange{\pageref{firstpage}--\pageref{lastpage}} \pubyear{2015}

\maketitle

\label{firstpage}

\begin{abstract}

Timing analysis provides information about the dynamics of matter accreting on to neutron stars and black holes, and hence is crucial for studying the physics of the accretion flow around these objects. It is difficult, however, to associate the different variability components with each of the spectral components of the accretion flow. We apply several new methods to two Rossi X-ray Timing Explorer observations of the black hole binary GRS 1915+105 during its heartbeat state to explore the origin of the X-ray variability and the interactions of the accretion-flow components. We offer a promising window into the disc--corona interaction through analysing the formation regions of the disc aperiodic variabilities with different time-scales via comparing the corresponding transition energies of the amplitude-ratio spectra. In a previous paper, we analysed the Fourier power density as a function of energy and frequency to study the origin of the aperiodic variability, and combined that analysis with the phase lag as a function of frequency to derive a picture of the disc--corona interaction in this source. We here, for the first time, investigate the phase lag as a function of energy and frequency, and display some interesting details of the disc--corona interaction. Besides, the results from the shape of amplitude-ratio spectrum and from several other aspects suggest that the quasi-periodic oscillation originates from the corona.

\end{abstract}

\begin{keywords}
accretion, accretion discs -- black hole physics -- X-rays: binaries -- X-rays: individual: GRS~1915+105
\end{keywords}

\section{INTRODUCTION}

More than 300 X-ray binaries have been detected \citep[e.g.][]{Liu2006, Liu2007}, among which more than 50 are black hole binaries \citep[BHBs;][]{Corral2016}. Accretion plays an important role in the formation and evolution of the X-ray binaries \citep[e.g.][]{Tauris2006, Zhang2011, Ji2011, Yan2012, Li2015, Weng2011, Weng2014}. There has been remarkable progress in understanding the black hole (BH) accretion flow， \citep[e.g.][]{Shakura73, Abramowicz88, Narayan95, Blandford99, Liu1999, Zhang2000, Yuan01, Falcke04, Liu2004, Meier05, Done07, Feng2011, Liu2013, Miller14, Yuan14, Gu2016, Wu2016}, however, many unresolved issues remain. The X-ray aperiodic variability and quasi-periodic oscillation (QPO) in the Fourier power density spectrum (PDS) are important tools to study the accretion flow. Nevertheless, it is hard to ascertain the origin of the X-ray variability due to the spectral overlap and the interactions of different accretion-flow components (e.g. accretion disc, and corona). For the aperiodic variability, it has been suggested that the high-frequency (more than several Hz) aperiodic variability produced in the accretion disc would be filtered out by the disc itself and the observed high-frequency aperiodic variability should come from the inner coronal flow \citep[e.g.][]{Lyubarskii97, Nowak99, Revnivtsev99, Klis2006, Titarchuk07, Gierlinski08, Wilkinson09, Heil11, Ingram11, Axelsson13}, while the power-law noise component should come from the accretion disc \citep[e.g.][]{Wilkinson09, Yu13, Stiele14}. 

The X-ray binary GRS 1915+105 contains a rapidly spinning BH \citep {Zhang97, McClintock06, Middleton06, Miller13} with a mass of $\sim 12$ $M_{\odot}$, and a K-M III giant star with a mass of $\sim 0.8$ $M_{\odot}$ as the donor \citep{Greiner01b, Harlaftis04, Reid14}. The orbital separation of the binary components is $\sim 108$ $R_{\odot}$ and the orbital period is $\sim 34$ days \citep{Greiner01a}. It is at a distance of $\sim 10$ kpc \citep[e.g.][]{Fender99, Zdziarski05, Reid14}, and is one of the strongest X-ray sources in the sky \citep{Remillard2006, McClintock2006}. A thermally driven disc wind has been detected in this source \citep{Lee2002, Ueda2009, Neilsen2009}.

During the heartbeat state \citep[named as the $\rho$ class by][]{Belloni00}, GRS 1915+105 oscillates quasi-periodically between a low-luminosity, spectrally hard state and a high-luminosity, soft state with periods less than $\sim 150$ s \citep{Neilsen12}. The fast state transitions and the high brightness make the $\rho$ class suitable for studying the X-ray variability.

There are several types of $\rho$ class observations with one to several pulses/peaks per cycle in the light curve \citep[e.g.][]{Taam97, Vilhu98, Belloni00, Massaro10}. We call the single-peaked $\rho$ class the $\rho_1$ class, and the double-peaked $\rho$ class the $\rho_2$ class. The spectral analysis performed by \citet{Neilsen11,Neilsen12} indicates that both classes have a similar accretion process, namely, the disc is a local Eddinton limit disc in the slow rising phase (the low/hard phase), and becomes unstable due to a thermal-viscous radiation pressure instability in the pulse phase (the high/soft phase). \citet{Massaro14} also suggests that the $\rho$ class can be modelled with a non-linear oscillator model containing two variables linked to the disc luminosity and temperature. \citet{Neilsen11,Neilsen12} argue that the single pulse in the $\rho_1$ class corresponds to the second, hard pulse in the $\rho_2$ class, and the difference in light curve results from the detailed shapes of the oscillations in the disc accretion rate. It is, therefore, interesting to compare the $\rho_1$ and $\rho_2$ classes through timing analysis.

We have presented the Fourier power density as a function of energy and frequency (the energy--frequency--power map) as a new method for investigating the origin of the X-ray variability, and first shown that the time-scale of the dominant variability of the corona is $\gtrsim$ 1 s \citep{Yan2017}, providing an important reference for studying the geometry and energy generation mechanism of the coronae in BHBs. Actually, the time lags between the soft X-ray bursts and the dips in the hard X-ray during type I bursts have been used to study the coronae in Neutron star X-ray binaries \citep[e.g.][]{Zhang2009, Chen2012, Ji2013}. 

We have also obtained a picture of the interaction between the disc and the corona in the $\rho$ class through a combination analysis of the spectral result, the energy--frequency--power map, and the frequency dependence of phase-lag \citep[the phase-lag spectrum;][]{Yan2017}. The energy dependence of the phase-lag should contain more comprehensive information of the interaction process than the phase-lag spectrum. Therefore, here we first plot the phase-lag as a function of energy and frequency (the energy--frequency--phase-lag map) to investigate the details of the disc--corona interaction. 

For one phase interval of the $\rho_2$ class, a high-frequency ($\gtrsim$ 10 Hz) disc aperiodic variability is first identified through a method which we call the amplitude-ratio spectrum (the energy dependence of the ratio of broad-band-noise to QPO amplitude) method \citep{Yan13apj}. The transition energy, $E_{\rm tr}$, of the amplitude-ratio spectrum is the energy below which the disc aperiodic variability becomes significant. It provides the information about the formation region of the disc aperiodic variability, which is also useful for analysing the disc--corona interaction. In this work, we thus apply the amplitude-ratio spectrum method to different $\rho$-cycle phases of the two $\rho$ classes in order to further investigate the disc--corona interaction.

Here, we analyse the QPO, the amplitude-ratio spectrum, and the energy--frequency--phase-lag map at different $\rho$ cycle phases in the $\rho_1$ and $\rho_2$ classes to explore the QPO origin, the disc--corona interaction, and compare the two classes. We describe the observations and data reduction methods in Section \ref{sec:data}, show the results in Section \ref{sec:result}, and present the discussion and the conclusions in Sections \ref{sec:discuss} and \ref{sec:con}, respectively.

\section{OBSERVATIONS AND DATA REDUCTION}\label{sec:data}

We have analysed two \emph{Rossi X-ray Timing Explorer} ({\it RXTE}) observations of GRS 1915+105. With the phase-folding method of \citet{Neilsen11}, we have performed a phase-resolved timing analysis of the $\rho_2$ class observation, obsid 60405-01-02-00 in \citet{Yan13apj}, and for the $\rho_1$ class observation, obsid 40703-01-07-00 in \citet{Yan2017}. 

In order to obtain a phase-folded {\it RXTE}/Proportional Counter Array (PCA) light curve, we extract two dead-time-corrected and background subtracted light curves with a time resolution of 1~s from the binned-mode data, B\_8ms\_16A\_0\_35\_H\_4P, covering the PCA energy channels 0 to 35, and the event-mode data, E\_16us\_16B\_36\_1s, covering the energy channels 36 to 249, respectively. We delete the asynchronous rows of the two light curves and add them together to obtain a light curve in the full PCA band to which we apply the barycenter correction. We then fold the barycentered light curve to obtain an average folded light curve. We determine the start time of each cycle by an iterative cross-correlation method \citep[see][]{Neilsen11}. We obtain 209 cycles with a mean period of 44.54 s for the $\rho_1$ class observation, and 257 cycles with a mean period of 50.29 s for the $\rho_2$ class observation.

For the timing analysis, we extract the light curves with a time resolution of 8 ms, calculate the PDS using 2 s segments, subtract the dead-time-corrected Poisson noise \citep{Morgan97}, and normalize the PDS to units of (rms/mean)$^2$/Hz \citep[e.g.][]{Miyamoto92}. We fit the PDS with a model including several Lorentzians to represent the broad-band-noise and the QPOs \citep{Nowak00, Belloni02}. For each 0.04 phase interval of the $\rho_1$ class, we calculate the PDS in the 1.9--38.4 keV band. For each 0.02 phase interval of the $\rho_2$ class, we calculate the PDS in the 2.1--37.8 keV band. The energy bands for the PDS calculations cannot be the same because the two $\rho$ classes were observed in different gain epochs while the data were archived with the same binning modes.

We extract the light curves in different energy bands to obtain the spectra of the amplitude of the broad-band-noise component, and of the QPO amplitude and frequency for several phases of the two $\rho$ classes. Since we aim to study the timing properties, we divide the cycle phase into phases I (0.00--0.08), II (0.08--0.26), III (0.26--0.40), IV (0.40--0.60), and 0.60--1.00 for the $\rho_1$ class observation, and phases i (0.02--0.12), ii (0.12--0.26), iii (0.26--0.40), iv (0.40--0.74), v (0.74--0.92), and vi (0.92--0.02) for the $\rho_2$ class observation based on the behaviour of the QPO amplitude (Fig. \ref{fig:lan.QPO}(c)). The phase interval 0.6--1.00 of the $\rho_1$ class is further divided into V (0.60--0.90) and VI (0.90--1.00) which roughly belong to the slow rising phase and the pulse phase, respectively \citep[Fig. \ref{fig:lan.QPO}(a);][]{Neilsen12}. We calculate the broad-band-noise amplitude by integrating the power in a frequency band, subtracting the contributions of the QPO and its harmonics \citep[e.g.][]{Vaughan03}, and correcting the amplitudes of the broad-band-noise and QPO for background \citep{Berger94, Rodriguez11}. We divide the broad-band-noise amplitude by the QPO amplitude to obtain the amplitude-ratio spectrum. 

We fit the energy dependence of QPO frequency (QPO frequency spectrum) and the amplitude-ratio spectrum using the Levenberg-Marquardt least-squares method implemented in {\scshape mpfit}, a programme written by \citet{Markwardt2009}. In addition, we use the Monte Carlo method to obtain the uncertainty ranges of the parameters in order to determine whether the least-squares fitting results are appropriate. We seed each spectrum $10^5$ times with randomly generated noise based on the standard deviations of the spectrum, and re-fit these simulated spectra. We calculate the 1$\sigma$ uncertainty range for each parameter based on their distribution obtained from the $10^5$ results, and fit the distribution with a Gaussian function to get each actual value. 

We utilize the phase-lag spectra to produce the energy--frequency--phase-lag maps for these two observations. The phase-lag spectra are obtained by calculating the Fourier cross-power spectra of two light curves extracted in different energy bands following \citet{Nowak99}. The selected energy bands are 1.94--4.05, 4.05--5.12, 5.12--7.25, 7.25--9.75, 9.75--12.99, 12.99--18.09, and 18.09--38.44 keV for the $\rho_1$ class observation, and  2.06--3.68, 3.68--5.31, 5.31--7.35, 7.35--9.81, 9.81--14.76, 14.76--18.10, 18.10--24.45, and 24.45--37.78 keV for the $\rho_2$ class observation. The reference energy band of the phase-lag is 5.12--7.25 keV for the $\rho_1$ class observation, and is 5.31--7.35 keV for the $\rho_2$ class observation. We choose these two energy bands as reference bands based on the energy--channel relation of the two gain epochs, the binning modes of the two observations, and the count rates of the two bands. 

A positive phase-lag denotes that the variability in an energy band lags that in the reference band. We call the phase-lag hard lag when it is positively correlated with energy, soft lag when it is anti-correlated with energy, and zero lag when it is not correlated with energy. All error bars in this paper correspond to the 1$\sigma$ confidence level.

\section{RESULTS}\label{sec:result}

We present in this section results of the phase-resolved timing analysis of observations of the $\rho_1$ and $\rho_2$ classes of GRS 1915+105, including the phase dependence of the QPO for the $\rho_1$ class, the QPO frequency spectra, the amplitude-ratio spectra, and the energy--frequency--phase-lag maps. For the aperiodic variability, we define three frequency bands, the low-frequency band ($\sim$ 0.5--1 Hz), the intermediate-frequency band ($\sim$ 1--10 Hz), and the high-frequency band ($\sim$ 10--30 Hz).

\subsection{Phase dependence of the QPO}

\begin{figure}
\centerline{\includegraphics[height=12cm,angle=0]{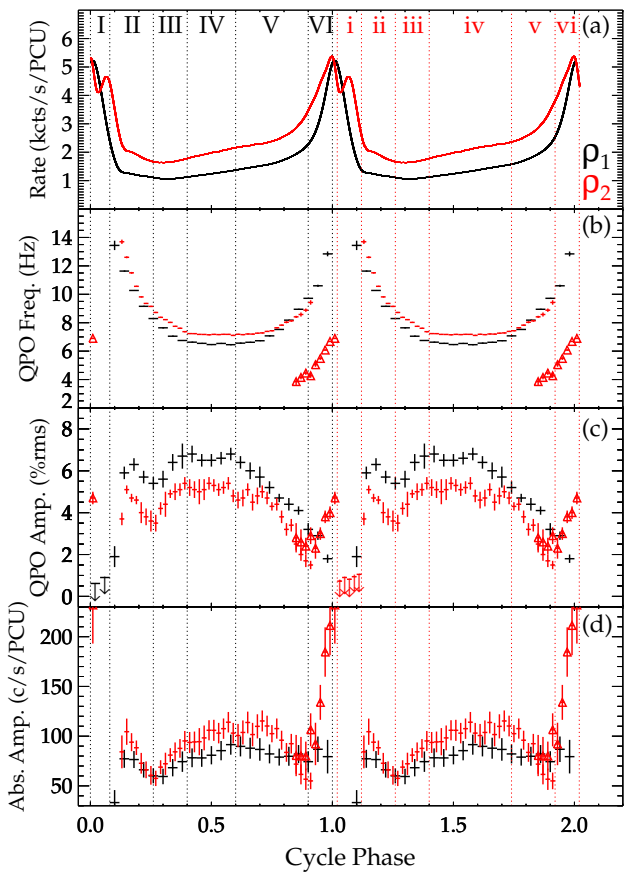}}
\caption{\label{fig:lan.QPO} The phase-folded PCA light curve in the 2--60 keV band (a), the QPO frequency (b), the QPO amplitude (c), and the QPO absolute amplitude (d) as a function of the $\rho$-cycle phase for the $\rho_1$ class (black; \emph{RXTE} Observation 40703-01-07-00) and the $\rho_2$ class (red; \emph{RXTE} Observation 60405-01-02-00) in GRS 1915+105. The red triangles represent the sub-harmonic QPO (SHQPO). The black and red vertical dotted lines and Roman numbers represent the phase intervals defined in Section \ref{sec:data} for the $\rho_1$ and $\rho_2$ classes, respectively. The upper limits of the QPO amplitude are indicated by downwards arrows.}
\end{figure}

The phase-folded RXTE/PCA light curve, and the phase dependence of the QPO frequency, amplitude, and absolute amplitude for the $\rho_1$ class are shown in black in Fig. \ref{fig:lan.QPO}. We present the corresponding results for the $\rho_2$ class in \citet{Yan13apj,Yan2017}, and overplot them here in red for comparison.

The QPO frequency evolves smoothly in the $\rho_1$ class. As the phase increases, the QPO frequency decreases (II and III), flattens (IV), and then increases again (V and VI; Fig. \ref{fig:lan.QPO}b). In the $\rho_2$ class, the QPO disappears at $\phi \gtrsim$ 0.90, and its sub-harmonic QPO (SHQPO) appears in the phase 0.84--1.02. The QPO frequencies at phase $\phi \lesssim (\gtrsim)$ 0.80 in the $\rho_1$ class are lower (higher) than those in the $\rho_2$ class. 

In phase I, there is no obvious QPO; we thus calculate the upper limit of the QPO amplitude (indicated by the downwards arrows). We first fit the PDS with a model including several Lorentzians, and then add an additional Lorentzian to the model based on the residuals. When the lower confidence limits for the parameters of the last Lorentzian added do not converge, we use the upper limits at the 1$\sigma$ confidence level to calculate the QPO amplitude, which is taken as the upper limit of the QPO amplitude. The upper limits are computed at frequencies 6.00 Hz ($\phi$: 0.00--0.04) and 8.01 Hz ($\phi$: 0.04--0.08) for the $\rho_1$ class, and 8.75 Hz ($\phi$: 0.02--0.04), 4.90 Hz ($\phi$: 0.04--0.06), 8.25 Hz ($\phi$: 0.06--0.08), 8.22 Hz ($\phi$: 0.08--1.00), and 8.21 Hz ($\phi$: 1.00--1.02) for the $\rho_2$ class. The fittings are performed through ISIS Version 1.6.2--30 \citep{Houck2000}.

The QPO amplitude is uncorrelated with the count rate in phases II, III, and IV, and is anti-correlated with the count rate in phases V and VI. The QPO amplitude and frequency are positively correlated in phase II, and anti-correlated in the other phases. In view of the overall situation, the QPO amplitude dips in phases II and III, remains almost constant in phase IV, and decreases in phases V and VI. For all of the phases except I and VI, the QPO amplitude is higher in the $\rho_1$ class than in the $\rho_2$ class.

We compute the QPO absolute amplitude via multiplying the QPO amplitude by the count rate in the energy band where the QPO amplitude is computed \citep[e.g.][]{Mendez97, Gilfanov03, Zdziarski05}. As the phase increases, the absolute amplitude of the $\rho_1$ class decreases in phase II, increases in phases III and IV, decreases slightly in the phase 0.60--0.80, and then remains a constant in the phase 0.80--1.00. The absolute amplitude of the $\rho_2$ class has a similar trend, but with some obvious differences. As the phase increases, the absolute amplitude of the $\rho_2$ class is higher than that of the $\rho_1$ class in most of the phases, but drops to near it at the end of phase II and below it at the end of phase V. The absolute amplitude of the SHQPO shows a rapid rise in the phase 0.84--1.02.

\begin{figure}
\centerline{\includegraphics[height=11cm,angle=0]{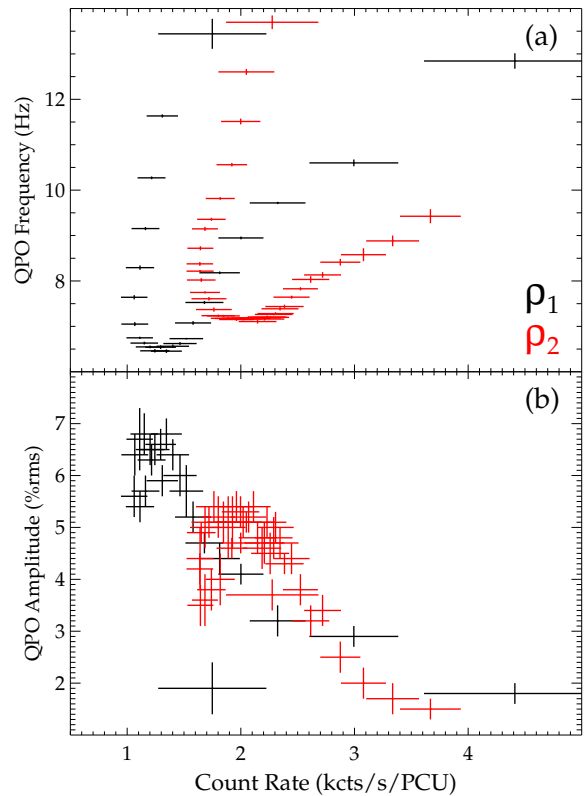}}
\caption{\label{fig:lanrate2fre2amp} The QPO frequency and amplitude as a function of the count rate of the $\rho_1$ (black) and $\rho_2$ class (red) in GRS 1915+105. The results of the SHQPO are not shown in this figure.}
\end{figure}

In order to compare the two $\rho$ classes more clearly, we present the count rate--QPO frequency and the count rate--QPO amplitude relations in Fig. \ref{fig:lanrate2fre2amp}. The shape of each relation in the $\rho_1$ class is similar to the corresponding one in the $\rho_2$ class, but with obvious shifts. The low frequency part of the count rate--QPO frequency relation (corresponding to phases IV and iv) shifts from low frequencies in the $\rho_1$ class to high frequencies in the $\rho_2$ class.

\begin{figure}
\centerline{\includegraphics[height=8.7cm,angle=-90]{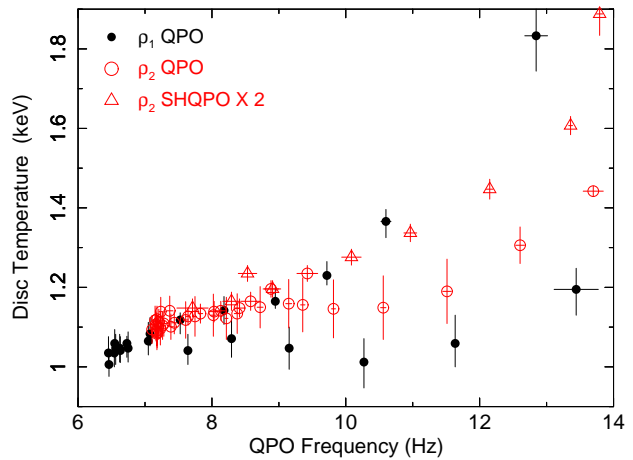}}
\caption{\label{fig:lan.f2t} The relations between the QPO frequency and the disc temperature in the $\rho_1$ (black pints) and $\rho_2$ class (red circles and triangles) in GRS 1915+105. The red circles correspond to QPO frequencies, while the red triangles correspond to the double of the SHQPO frequencies. The data of the disc temperature is from \citet{Neilsen11, Neilsen12}.}
\end{figure}

We present the relation between the QPO frequency and the disc temperature obtained from \citet{Neilsen11, Neilsen12} in Fig. \ref{fig:lan.f2t}. The relation shows a horizontal ``v'' shape in the $\rho_1$ class. For the $\rho_2$ class, the QPO frequency, along with the double of the SHQPO frequency, form a similar ``v'' shape relation with the disc temperature.

\subsection{QPO frequency spectra}\label{sec:frespectra}

\begin{figure}
\centerline{\includegraphics[height=9.55cm,angle=0]{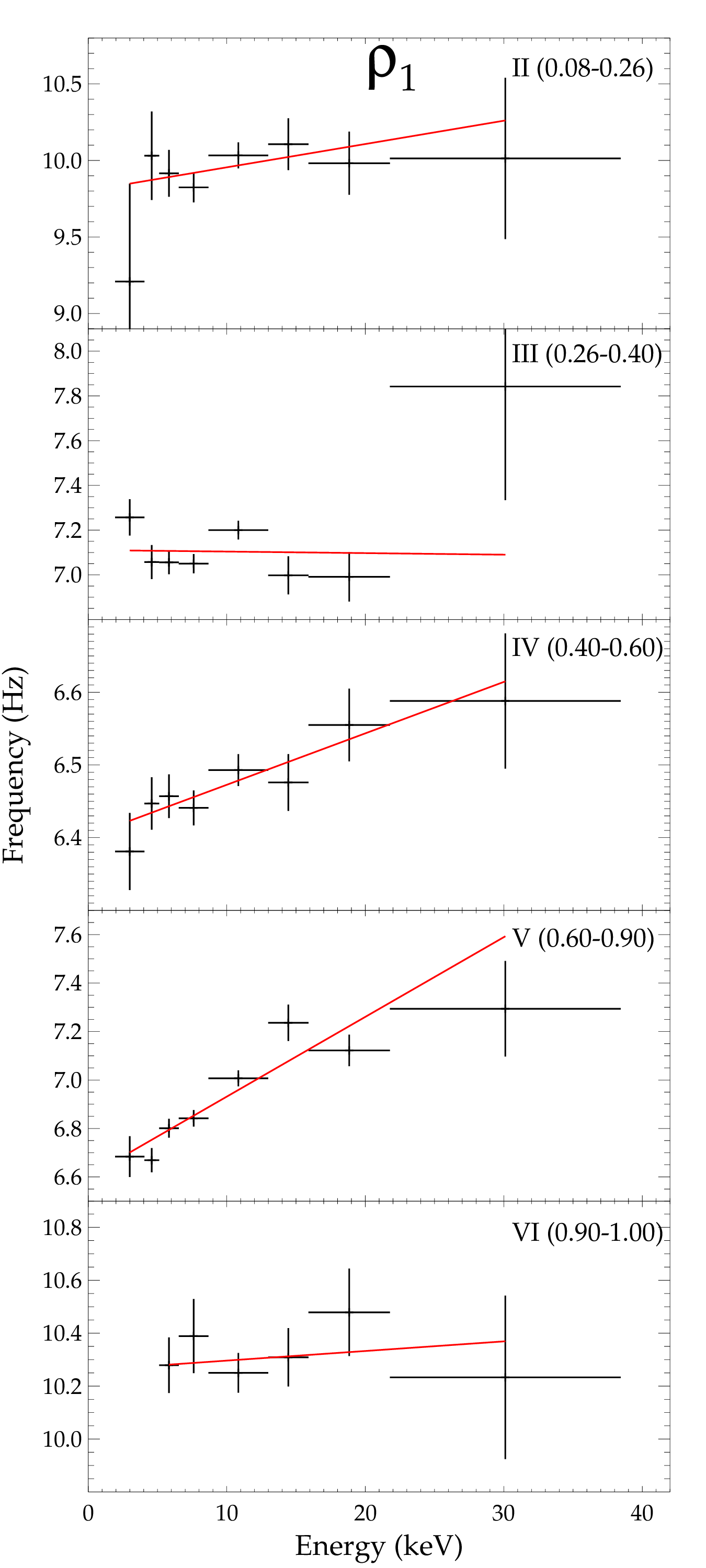}  \includegraphics[height=9.55cm,angle=0]{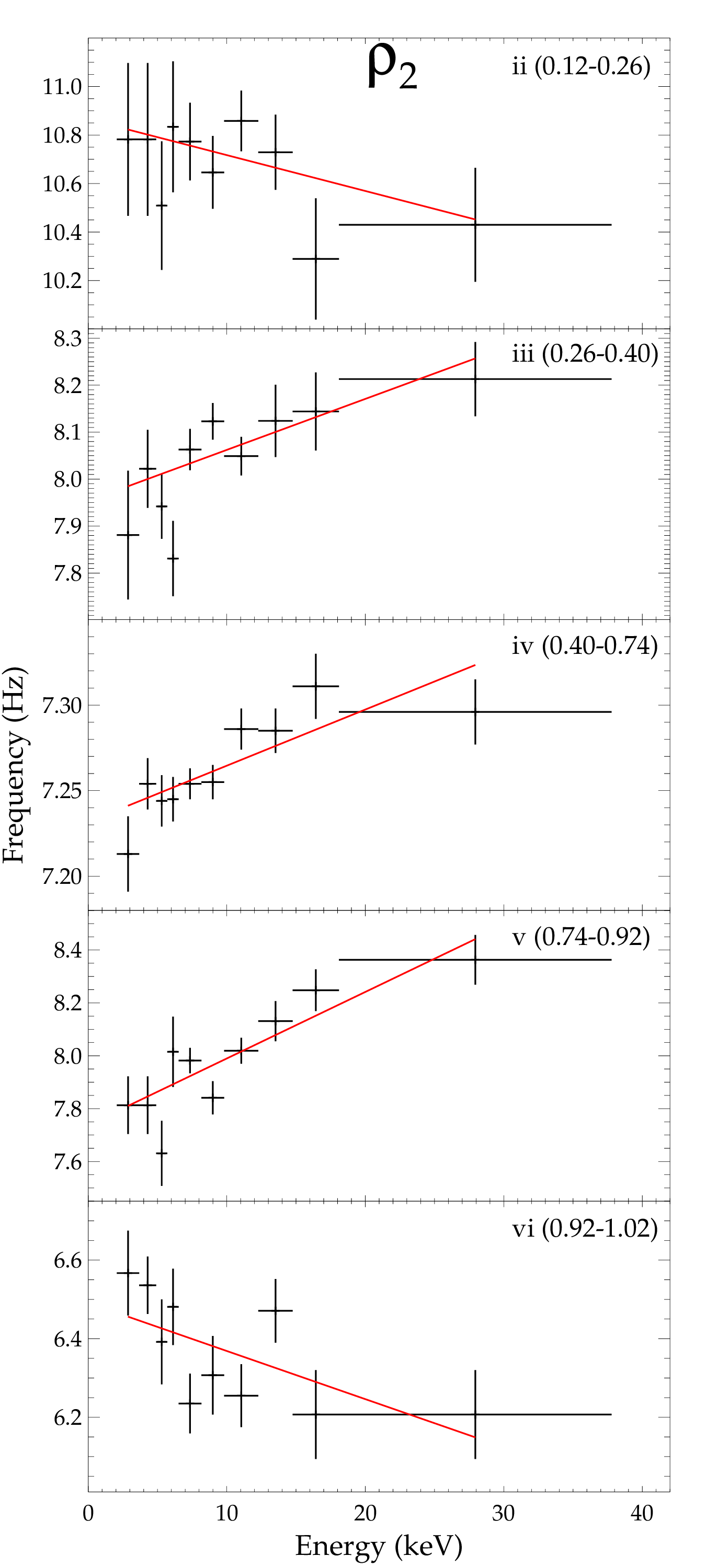}}
\caption{\label{fig:40703fre} The QPO frequency spectra of some phases of the $\rho_1$ and $\rho_2$ classes in GRS 1915+105. In order to obtain as many light curves with sufficient signal-to-noise ratio for timing analysis as possible, we select the energy bands 1.94--4.05, 4.05--5.12, 5.12--6.54, 6.54--8.68, 8.68--12.99, 12.99--15.90, 15.90--21.78, and 21.78--38.44 keV for the $\rho_1$ class observation, and 2.06--3.68, 3.68--4.90, 4.90--5.71, 5.71--6.53, 6.53--8.17, 8.17--9.81, 9.81--12.28, 12.28--14.76, 14.76--18.10, and 18.10--37.78 keV for the $\rho_2$ class observation. The plot for phase vi shows the result of the SHQPO of the $\rho_2$ class. The red lines are the best-fitting lines.}
\end{figure}

Fig. \ref{fig:40703fre} shows the QPO frequency spectra of the $\rho_1$ and $\rho_2$ classes. We fit the QPO frequency spectra with linear functions (red lines) with {\scshape mpfit}, and show slopes of the best-fitting linear functions in Table \ref{table1}, and then perform Monte Carlo simulations to test whether the slopes are credible. We find that the fitting results are well in agreement with the simulated results, and therefore not show it in Table \ref{table1}. We also calculate the weighted Pearson correlation coefficients for the data in each panel. The Pearson correlation coefficient goes from 1 if there is a perfect positive linear correlation to $-1$ if there is a perfect anti-correlation \footnote{\url{https://en.wikipedia.org/wiki/Pearson_correlation_coefficient}}. Besides, in order to assess the significance of the dependence of the QPO frequency on energy, we fit both a constant and a linear function to each QPO frequency spectrum and calculate the F-test probability.

The results show different behaviour between the slow rising phase (including phases III, IV, V of the $\rho_1$ class, and iii, iv, v of the $\rho_2$ class) and the pulse and post-pulse phases (including phases VI, vi, as well as II and ii). 

In phases IV, V and iii, iv, v, the F-test probabilities range 0.037 to 0.001, indicating that a linear function is significantly better than a constant for the QPO frequency spectra in these phases. And the weighted Pearson correlation coefficients are between 0.66 and 0.91, suggesting that the QPO frequency is positively correlated with energy. The slope value in phase III is around zero, however the QPO frequency is significantly higher in the range 20--40 keV than in the range 2--20 keV band. 

On the contrary, the results in the pulse and post-pulse phases are obviously different with respect to the slow rising phase: the QPO frequency almost does not vary with the energy (phases II, VI, and ii; the F-test probabilities range 0.632 to 0.123 and the correlation coefficients are between $-$0.52 to 0.53) or even decreases with the energy (phase vi; the F-test probability is 0.086 and the correlation coefficient is $-$0.57).

\begin{table*}
\centering
\caption{The fitting results, the weighted Pearson correlation coefficients, and the F-test probabilities for the QPO frequency spectra of some phases of the $\rho_1$ and $\rho_2$ classes in GRS 1915+105. The row for phase vi shows the results of the SHQPO of the $\rho_2$ class.}
\begin{minipage}{10cm}

\label{table1}
\begin{tabular}{|c|c|cc|c|c|}
\hline
\multirow{2}{*}{Class} & \multirow{2}{*}{Phase} & \multicolumn{2}{c|}{Linear fitting} & \multirow{2}{*}{\parbox{0.5in}{\centering{Pearson\\correlation\\coefficient}}} & \multirow{2}{*}{\parbox{0.52in}{\centering{F-test\\Probability}}}\\
\cline{3-4}
 &  & Slope (Hz/keV) & Reduced $\chi^2$ & & \\
\hline
\multirow{5}{*}{\rotatebox{0}{$\rho_1$}}
& II (0.08--0.26)  & $0.015 \pm 0.013$ & 0.60 & 0.53 & 0.182\\
& III (0.26--0.40) & $-0.001 \pm 0.006$ & 2.71 & $-0.03$ & 0.954\\
& IV (0.40--0.60)  & $0.007 \pm 0.003$ & 0.42 & 0.87 & 0.005\\
& V (0.60--0.90)  & $0.033 \pm 0.004$ & 2.38 & 0.91 & 0.002\\
& VI (0.90--1.00)  & $0.004 \pm 0.010$ & 0.50 & 0.25 & 0.632\\
\hline
\multirow{5}{*}{\rotatebox{0}{$\rho_2$}}
& ii (0.12--0.26)  & $-0.015 \pm 0.011$ & 0.63 & $-0.52$ & 0.123\\
& iii (0.26--0.40) & $0.011 \pm 0.004$ & 1.48 & 0.66 & 0.037\\
& iv (0.40--0.74) & $0.003 \pm 0.001$ & 1.17 & 0.81 & 0.005\\
& v (0.74--0.92)  & $0.025 \pm 0.004$ & 1.60 & 0.86 & 0.001\\
& vi (0.92--0.02)  & $-0.012 \pm 0.005$ & 1.79 & $-0.57$ & 0.086\\
\hline

\end{tabular}
\end{minipage}
\medskip
\end{table*}

\subsection{Amplitude-ratio spectra}\label{sec:ratio}

\begin{figure*}
\centering
\subfigure{
\begin{minipage}[b]{1.0\textwidth}
\includegraphics[width=1\textwidth]{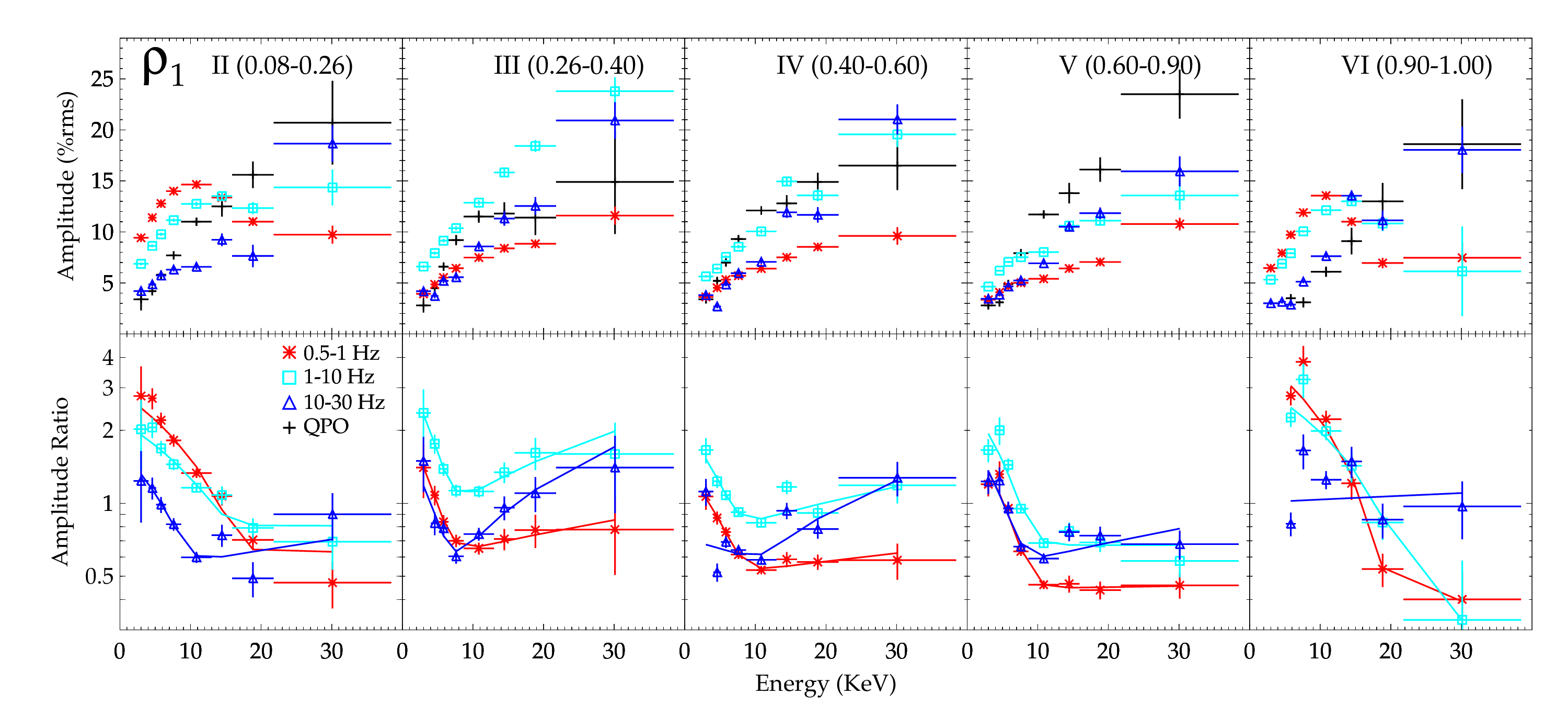}
\end{minipage}
}
\subfigure{
\begin{minipage}[b]{1.0\textwidth}
\includegraphics[width=1\textwidth]{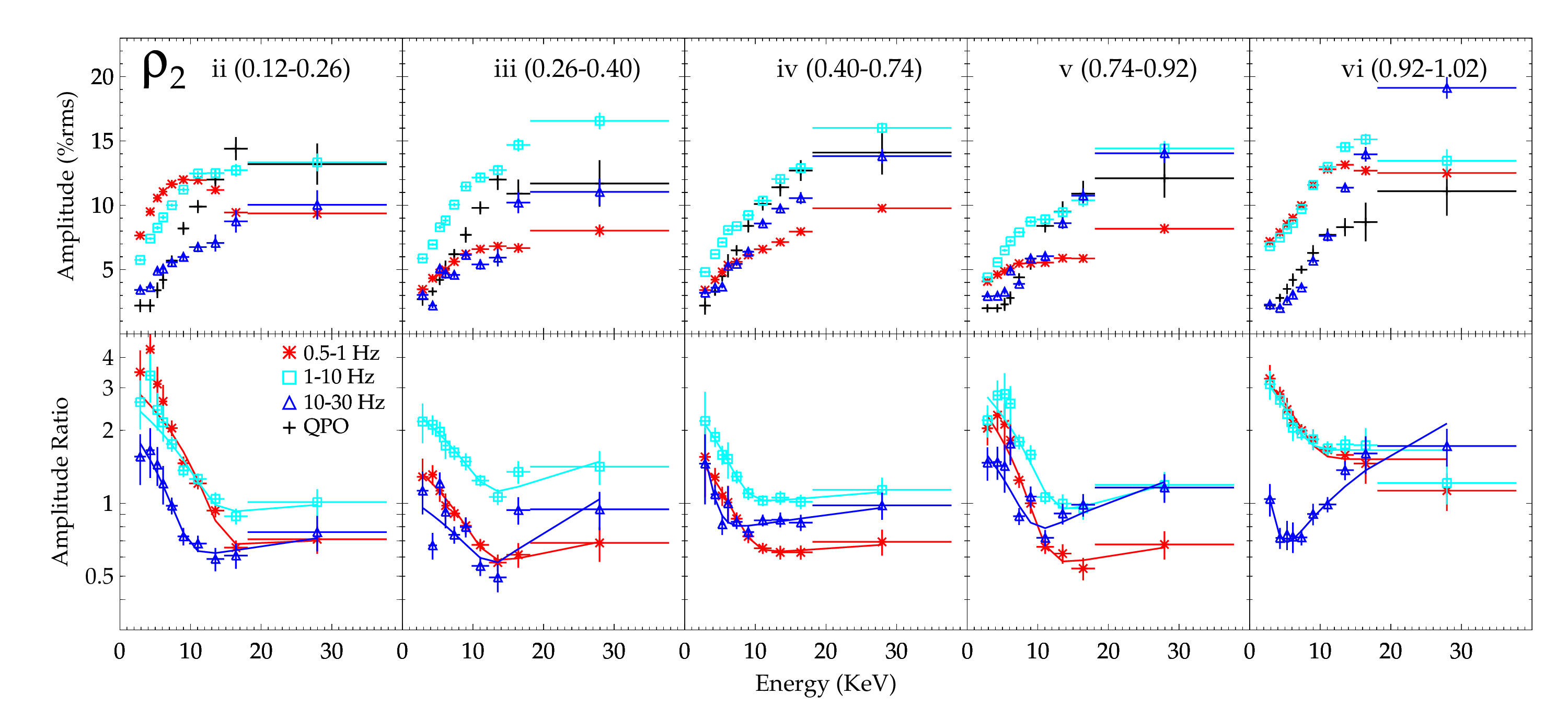}
\end{minipage}
}
\caption{\label{fig:rmsratio} The energy dependence of the broad-band-noise amplitudes in three frequency ranges (0.5--1 Hz, 1--10 Hz, and 10--30 Hz) and of the QPO amplitude (the upper row of each subgraph), as well as the energy dependence of the ratio of broad-band-noise to QPO amplitude (the lower row of each subgraph) of some phases of the $\rho_1$ and $\rho_2$ classes in GRS 1915+105. The plots for phase vi show the results of the SHQPO of the $\rho_2$ class.} The red asterisks, the cyan squares, and the blue triangles represent the 0.5--1 Hz, 1--10 Hz, and 10--30 Hz broad-band-noise amplitude spectra and the corresponding amplitude-ratio spectra, respectively. The black cross represent the QPO amplitude spectra. The lines are the best-fitting lines.
\end{figure*}

We illustrate the broad-band-noise amplitude spectra and the QPO amplitude spectra of the $\rho_1$ and $\rho_2$ classes in the upper panels of each subgraph of Fig. \ref{fig:rmsratio}, and show the amplitude-ratio spectra in the lower panels of each subgraph of Fig. \ref{fig:rmsratio}. In phases II, VI of the $\rho_1$ class and ii, vi of the $\rho_2$ class, the 0.5--1 Hz amplitude spectrum shows a bump at $\sim$ 10 keV, and the 1--10 Hz amplitude spectrum shows a bump at $\sim$ 15 keV. The other broad-band-noises and all of the QPOs show a positive correlation between amplitude and energy. 

\begin{figure}
\centerline{\includegraphics[height=9.6cm,angle=0]{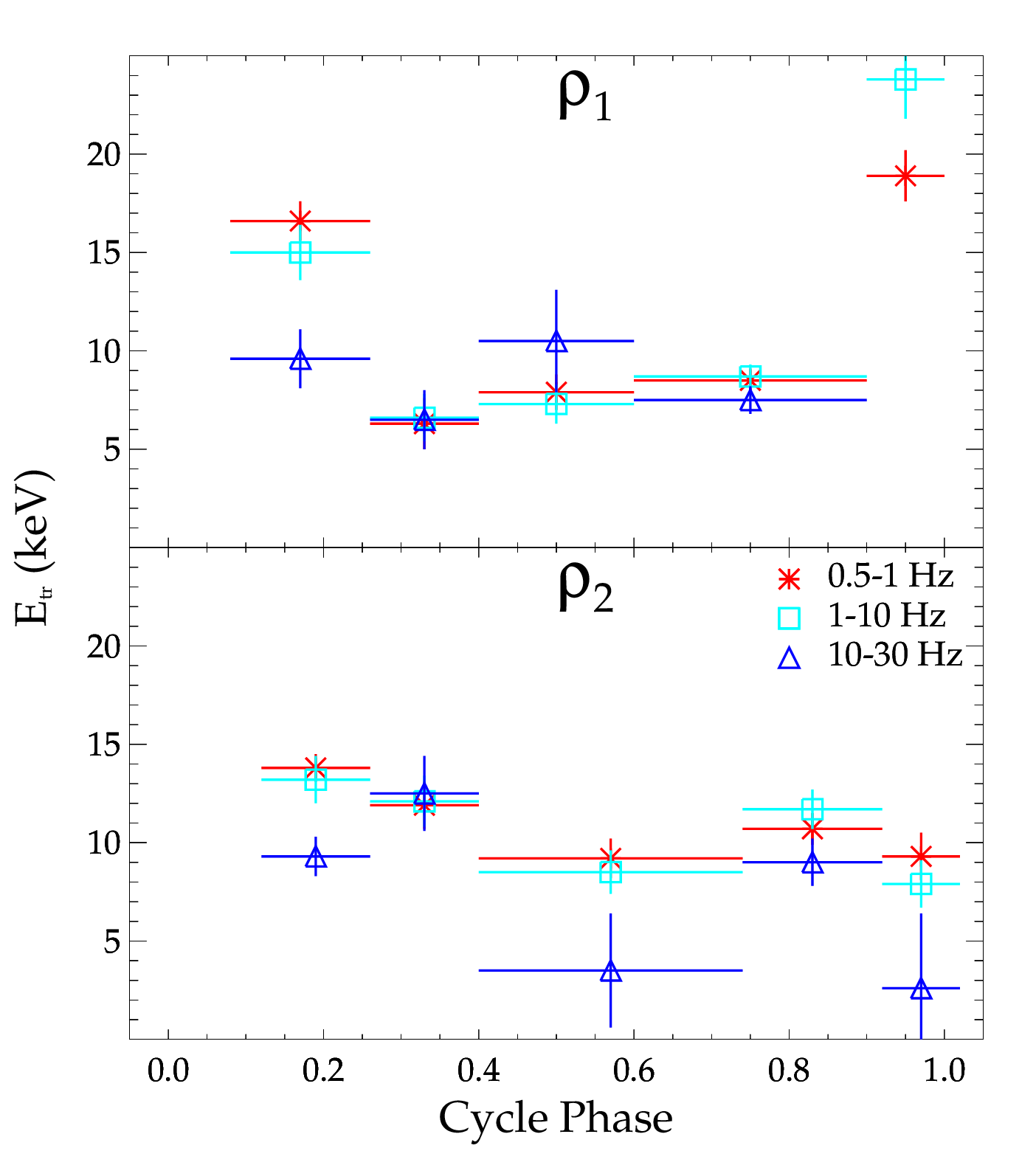}}
\caption{\label{fig:p2e} The relation between the transition energy and the cycle phase for the $\rho_1$ and $\rho_2$ classes in GRS 1915+105. The red asterisks, the cyan squares, and the blue triangles represent the transition energies of the 0.5--1 Hz, 1--10 Hz, and 10--30 Hz amplitude-ratio spectra, respectively.}
\end{figure}

All of the broad-band-noise to QPO amplitude-ratios, except the 10--30 Hz one in phase VI of the $\rho_1$ class, decrease with energy at low energies, and then smoothly level off or increase with energy at high energies. This behaviour can be described by the function $R(E)=C-S_{1}\ln(\exp (E_{\rm tr}-E)+1)+S_{2}E$, where $E_{\rm tr}$, $S_{1}$, $S_{2}$, and $C$ are the parameters. The two asymptotes of this function are $R(E)=(S_{1}+S_{2})E+(C-S_{1}E_{\rm tr})$ for $E < E_{\rm tr}$ and $R(E)=S_{2}E+C$ for $E > E_{\rm tr}$. In this work, $|S_{2}| \ll |S_{1}|$. Thus, $S_{1}$ is the slope of the low-energy part of the amplitude-ratio, $S_{2} \geq 0$ is the slope of the high-energy part of the amplitude-ratio, $E_{\rm tr}$ is the energy at which the transition of the amplitude-ratio occurs, and $C \geq 0$ is a constant term. We fit the 10--30 Hz amplitude-ratio spectrum in phase VI of the $\rho_1$ class with a linear function. The least-squares fitting results are consistent with the Monte Carlo simulated results for the amplitude-ratio spectra in all phases of the two $\rho$ class, except phase IV of the $\rho_1$ class (Table \ref{table2}; see also the following paragraph). We present the relations between the $E_{\rm tr}$ and the phase for the two classes in Fig. \ref{fig:p2e}.

For the $\rho_1$ class, $E_{\rm tr}$ of the 0.5--1 Hz amplitude-ratio spectrum is close to that of the 1--10 Hz amplitude-ratio spectrum for all of the phases, $E_{\rm tr}$ of the 10--30 Hz amplitude-ratio spectrum is also similar to that of the 0.5--1 Hz and 1--10 Hz amplitude-ratio spectra for the slow rising phase (III, IV, and V), but lower than them for the pulse phase (VI) and post-pulse phase (II) (see Table \ref{table2} and Fig. \ref{fig:p2e}). For phase IV of the $\rho_1$ class, $E_{\rm tr}$ of the 10--30 Hz amplitude-ratio spectrum is higher than that of the 0.5--1 Hz and 1--10 Hz amplitude-ratio spectra based on the results of the Monte Carlo simulations. However, $E_{\rm tr}$ of these three amplitude-ratio spectra are similar when checked by eye or based on the results of the least-squares fittings.

For the $\rho_2$ class, $E_{\rm tr}$ of the 0.5--1 Hz amplitude-ratio spectrum is similar to that of the 1--10 Hz amplitude-ratio spectrum for all of the phases, $E_{\rm tr}$ of the 10--30 Hz amplitude-ratio spectrum is also similar to that of the 0.5--1 Hz and 1--10 Hz amplitude-ratio spectra for part of the slow rising phase (iii and v), but lower than them for the pulse phase (vi) and post-pulse phase (ii). While for the central part of the slow rising phase (iv), $E_{\rm tr}$ of the 10--30 Hz amplitude-ratio spectrum is also lower than that of the 0.5--1 Hz and 1--10 Hz amplitude-ratio spectra.

\begin{table*}
\centering
\begin{minipage}{17.5cm}
\caption{Least-squares fittings and Monte Carlo Simulations for the amplitude-ratio spectra of some phases of the $\rho_1$ and $\rho_2$ classes in GRS 1915+105. The rows for phase vi show the results of the SHQPO of the $\rho_2$ class.}
\label{table2}
\begin{tabular}{|c|c|c|ccccc|c|}

\hline
\multirow{2}{*}{Class} & \multirow{2}{*}{Phase} & \multirow{2}{*}{broad-band-noise} &\multicolumn{5}{c|}{$R(E)=C-S_{1} \ln(\exp(E_{\rm tr}-E)+1)+S_{2}E$} &\multicolumn{1}{c|}{Monte Carlo Sim.} \\
\cline{4-9}
& & & $E_{\rm tr}$ (KeV) & $S_{1}$ (KeV$^{-1}$) & $S_{2}$ (KeV$^{-1}$) & $C$ & reduced $\chi^2$ & $E_{\rm tr}$ (KeV) \\
\hline
\multirow{15}{*}{\rotatebox{0}{$\rho_1$}}&\multirow{3}{*}{II(0.08--0.26)}
  &0.5--1 Hz& $16.6 \pm 1.0$& $-0.14 \pm 0.02$&        $0$       & $0.63 \pm 0.06$& 2.20& $16.2^{+2.0}_{-3.1}$\\
& & 1--10 Hz& $15.0 \pm 1.4$& $-0.09 \pm 0.02$&        $0$       & $0.81 \pm 0.07$& 1.49& $14.2^{+2.9}_{-2.8}$\\
& &10--30 Hz&  $9.6 \pm 1.5$& $-0.12 \pm 0.04$& $0.007 \pm 0.009$& $0.50 \pm 0.14$& 1.83& $9.6^{+1.2}_{-1.2}$\\
\cline{2-9}
&\multirow{3}{*}{III(0.26--0.40)}
  &0.5--1 Hz&  $6.3 \pm 1.3$& $-0.25 \pm 0.15$& $0.010 \pm 0.009$& $0.55 \pm 0.12$& 0.11& $6.4^{+1.5}_{-1.5}$\\
& & 1--10 Hz&  $6.6 \pm 1.2$& $-0.43 \pm 0.21$& $0.044 \pm 0.018$& $0.65 \pm 0.22$& 0.26& $6.6^{+1.3}_{-1.3}$\\
& & 10--30 Hz& $6.5 \pm 1.5$& $-0.24 \pm 0.14$& $0.051 \pm 0.014$& $0.18 \pm 0.17$& 0.87& $6.7^{+1.3}_{-1.2}$\\
\cline{2-9}
&\multirow{3}{*}{IV(0.40--0.60)}
  &0.5--1 Hz&  $7.9 \pm 0.9$& $-0.11 \pm 0.03$& $0.005 \pm 0.004$& $0.48 \pm 0.06$& 0.30& $7.9^{+0.8}_{-0.8}$\\
& & 1--10 Hz&  $7.3 \pm 1.0$& $-0.18 \pm 0.06$& $0.016 \pm 0.007$& $0.68 \pm 0.10$& 3.29& $7.3^{+1.3}_{-1.2}$\\
& & 10--30 Hz&$10.5 \pm 2.6$& $-0.05 \pm 0.02$& $0.034 \pm 0.010$& $0.22 \pm 0.16$& 8.34& $10.5^{+0.7}_{-0.7}$\\
\cline{2-9}
&\multirow{3}{*}{V(0.60--0.90)}
  &0.5--1 Hz&  $8.5 \pm 0.6$& $-0.17 \pm 0.03$& $0.000 \pm 0.003$& $0.44 \pm 0.05$& 1.17& $8.5^{+0.5}_{-0.5}$\\
& & 1--10 Hz&  $8.7 \pm 0.6$& $-0.22 \pm 0.04$&        $0$       & $0.67 \pm 0.03$& 2.20& $8.6^{+0.4}_{-0.5}$\\
& & 10--30 Hz& $7.5 \pm 0.7$& $-0.19 \pm 0.05$& $0.010 \pm 0.005$& $0.49 \pm 0.06$& 2.43& $7.4^{+0.5}_{-0.5}$\\
\cline{2-9}
&\multirow{3}{*}{VI(0.90--1.00)}
  &0.5--1 Hz& $19.5 \pm 4.6$& $-0.21 \pm 0.04$& $0.008 \pm 0.063$& $0.15 \pm 1.87$& 2.99& $18.9^{+1.2}_{-1.1}$\\
& & 1--10 Hz& $23.8 \pm 2.0$& $-0.13 \pm 0.02$& $0.011 \pm 0.008$&       $0$      & 1.82& $23.1^{+2.3}_{-2.6}$\\
& & 10--30 Hz&      --      &        --       & $0.003 \pm 0.009$& $1.00 \pm 0.12$& 5.11&         --          \\
\hline
\multirow{15}{*}{\rotatebox{0}{$\rho_2$}}&\multirow{3}{*}{ii(0.12--0.26)}
  &0.5--1 Hz& $14.1 \pm 0.8$& $-0.20 \pm 0.03$& $0.003 \pm 0.009$& $0.60 \pm 0.19$& 2.20& $13.8^{+1.3}_{-1.2}$\\
& & 1--10 Hz& $13.2 \pm 1.2$& $-0.15 \pm 0.03$& $0.006 \pm 0.012$& $0.83 \pm 0.24$& 1.32& $13.3^{+2.8}_{-2.3}$\\
& & 10--30 Hz& $9.3 \pm 1.0$& $-0.19 \pm 0.06$& $0.007 \pm 0.009$& $0.53 \pm 0.14$& 0.37& $9.1^{+1.2}_{-1.2}$\\
\cline{2-9}
&\multirow{3}{*}{iii(0.26--0.40)}
  &0.5--1 Hz& $11.9 \pm 1.2$& $-0.09 \pm 0.02$& $0.009 \pm 0.009$& $0.45 \pm 0.16$& 0.30& $12.0^{+1.5}_{-1.5}$\\
& & 1--10 Hz& $12.1 \pm 1.3$& $-0.16 \pm 0.03$& $0.028 \pm 0.018$& $0.71 \pm 0.31$& 0.47& $12.1^{+1.2}_{-1.1}$\\
& & 10--30 Hz&$12.5 \pm 1.9$& $-0.08 \pm 0.02$& $0.034 \pm 0.014$& $0.08 \pm 0.26$& 4.51& $12.5^{+0.7}_{-0.8}$\\
\cline{2-9}
&\multirow{3}{*}{iv(0.40--0.74)}
  &0.5--1 Hz&  $9.2 \pm 1.0$& $-0.13 \pm 0.03$& $0.003 \pm 0.006$& $0.59 \pm 0.09$& 0.11& $9.3^{+1.2}_{-1.3}$\\
& & 1--10 Hz&  $8.5 \pm 1.1$& $-0.21 \pm 0.07$& $0.006 \pm 0.009$& $0.95 \pm 0.13$& 0.11& $8.6^{+1.2}_{-1.2}$\\
& & 10--30 Hz& $3.5 \pm 2.9$& $-0.74 \pm 1.65$& $0.008 \pm 0.006$& $0.73 \pm 0.07$& 0.62& $3.8^{+1.8}_{-2.0}$\\
\cline{2-9}
&\multirow{3}{*}{v(0.74--0.92)}
  &0.5--1 Hz& $10.7 \pm 0.8$& $-0.23 \pm 0.04$& $0.007 \pm 0.007$& $0.47 \pm 0.13$& 0.78& $10.7^{+0.8}_{-0.8}$\\
& & 1--10 Hz& $11.7 \pm 1.0$& $-0.23 \pm 0.04$& $0.019 \pm 0.013$& $0.65 \pm 0.25$& 1.19& $11.6^{+0.8}_{-0.7}$\\
& & 10--30 Hz& $9.0 \pm 1.2$& $-0.17 \pm 0.05$& $0.027 \pm 0.010$& $0.47 \pm 0.15$& 2.33& $9.0^{+1.3}_{-1.3}$\\
\cline{2-9}
&\multirow{3}{*}{vi(0.92--1.02)}
  &0.5--1 Hz&  $9.3 \pm 1.2$& $-0.25 \pm 0.07$&        $0$       & $1.52 \pm 0.08$& 0.89& $8.6^{+3.4}_{-2.2}$\\
& & 1--10 Hz&  $7.9 \pm 1.2$& $-0.28 \pm 0.10$&        $0$       & $1.66 \pm 0.08$& 0.74& $7.6^{+1.8}_{-2.0}$\\
& & 10--30 Hz& $2.6 \pm 3.8$& $-1.04 \pm 3.12$& $0.067 \pm 0.011$& $0.27 \pm 0.11$& 1.06& $4.0^{+1.7}_{-1.5}$\\
\hline

\end{tabular}
\label{table2}
\medskip
\end{minipage}
\end{table*}

\subsection{Energy--frequency--phase-lag maps}\label{sec:maps}

\begin{figure*}
\centerline{\includegraphics[height=19cm,angle=0]{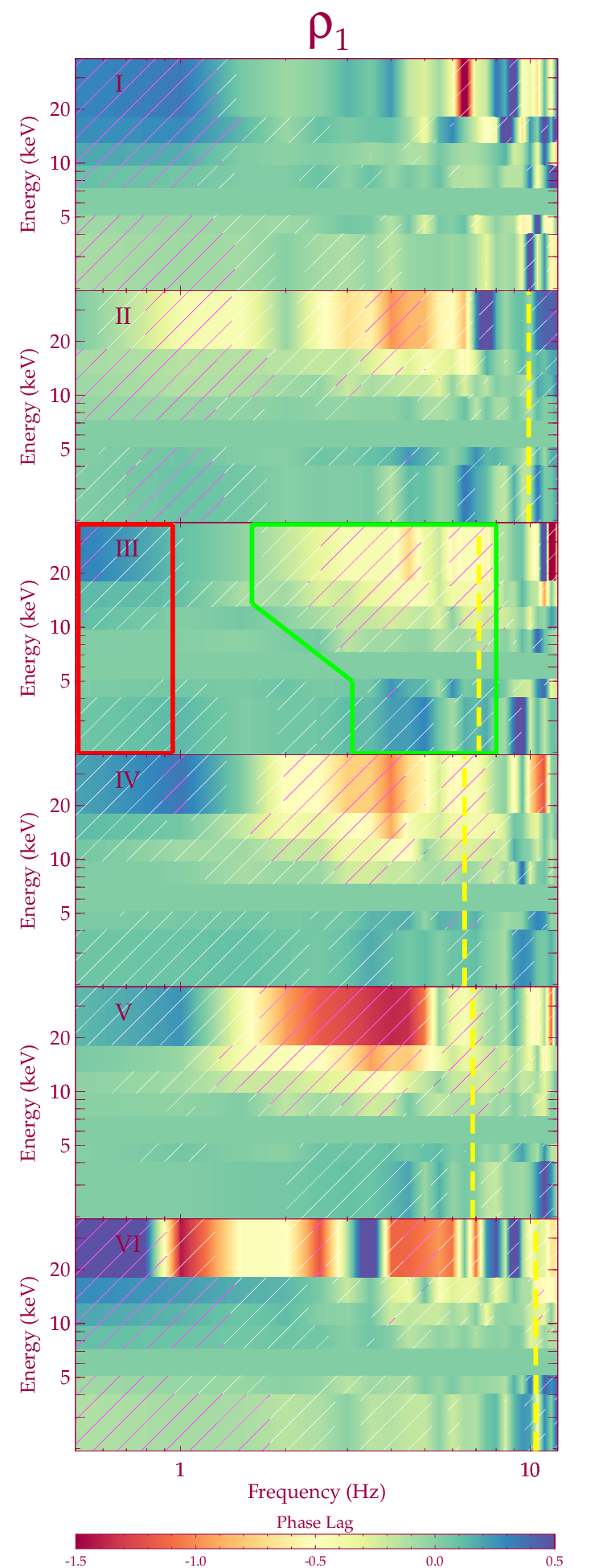}  \includegraphics[height=19cm,angle=0]{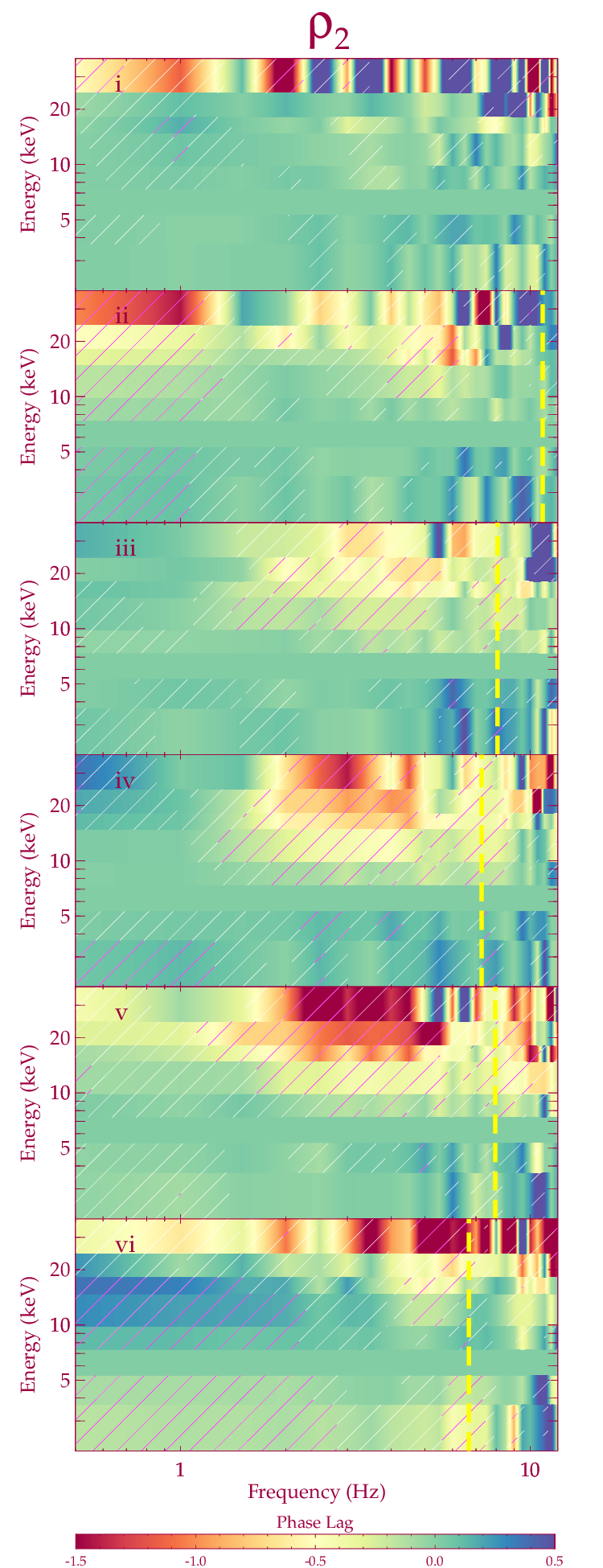}}
\caption{\label{fig:FEL40703} The energy--frequency--phase-lag maps of some phases of the $\rho_1$ and $\rho_2$ classes in GRS 1915+105. The plot for phase vi shows the result of the SHQPO of the $\rho_2$ class. The vertical yellow dashed lines denote the centroid frequency of the QPO. The red and green block diagrams for phase III of the $\rho_1$ class are used to denote the hard- and soft-lag regions, respectively, as an example for highlighting the phase-lag distribution. The magenta oblique lines indicate the regions where the phase-lags are more than 3$\sigma$ different from zero. The white oblique lines indicate the regions where the phase-lags are between 1$\sigma$ to 3$\sigma$ different from zero. The rest of the phase-lags, except those in the reference bands, are less than 1$\sigma$. The phase-lags higher than 0.5 and lower than -1.5 are set to blue and red, respectively.}
\end{figure*}

The energy--frequency--phase-lag maps show the phase-lag in the low- and intermediate-frequency bands as a function of energy and frequency (Fig. \ref{fig:FEL40703}). In the high-frequency band, above 12 Hz, the phase-lag distribution is not clear due to the low signal to noise ratio, we thus only show the phase-lag distribution in the 0.5--12 Hz band. As an example for highlighting the phase-lag distribution, we use the red and green block diagrams to, respectively, denote the hard- and soft-lag regions for phase III of the $\rho_1$ class. We compute the ratio between the phase-lag and its standard deviation to represent the significance of the detection of the phase-lag, and indicate the significance distribution in the same figure. The significance of the phase-lags covered by magenta oblique lines is higher than 3$\sigma$, the significance of those covered by white oblique lines is between 1$\sigma$ to 3$\sigma$, and the significance of the rest of the phase-lags, except those in the reference bands, is less than 1$\sigma$.

In phase I of the $\rho_1$ class, the phase-lag is hard in the low-frequency band, and approximates zero in the intermediate-frequency band. In phase II, the phase-lag is significantly soft in the low- and intermediate-frequency bands, except the lower-left corner of the plot. In phases III, IV, and V, the phase-lag is hard in the low-frequency band, and is soft in the intermediate-frequency band, except in the lower-left corner. In phase VI, the phase-lag distribution is similar to that of the previous three phase intervals, but the frequency range of the hard lag extends to above 2 Hz.

The phase-lag distributions of the two $\rho$ class are similar, but there is still a significant difference. The difference is that the hard lag is smaller in phases iii, iv, and v of the $\rho_2$ class than in phases III, IV, and V of the $\rho_1$ class, while it is slightly higher in phase vi of the $\rho_2$ class than in phase VI of the $\rho_1$ class, and approximates zero lag in phase i of the $\rho_2$ class.

The significance of the phase-lag detection is usually higher in the low-frequency band than in the intermediate-frequency band for phase I, II, and VI of the $\rho_1$ class and i, ii, and vi of the $\rho_2$ class, while lower in the low-frequency band than in the intermediate-frequency band for the rest of the phases.

\section{DISCUSSION}\label{sec:discuss}

We systematically study the QPO of GRS 1915+105 during the $\rho_1$ and $\rho_2$ classes from many aspects, including the phase dependence of the QPO, the QPO frequency and amplitude as a function of count rate, the relation between QPO frequency and disc temperature, the dependence of the QPO frequency on energy, and the shape of amplitude-ratio spectrum. We, for the first time, compare the transition energies of the low- and high-frequency amplitude-ratio spectra to explore the disc--corona interaction. Besides, we display the phase-lag as a function of energy and frequency as a new method to investigate the details of the disc--corona interaction.

In this section, we will discuss these results to analyse the origin of the QPO and the disc--corona interaction of GRS 1915+105 in heartbeat state.

\subsection{QPO's origin}\label{sec:discuss1}

The disc and the corona are two essential spectral components for GRS 1915+105 during the $\rho$ class \citep[e.g.][]{Neilsen11, Neilsen12, Mineo2012, Mineo2017}. Figs. \ref{fig:lan.f2t} and \ref{fig:lanrate2fre2amp} show that the QPO frequency is correlated with the disc temperature and with the flux of the total emission that is dominated by the disc \citep{Neilsen11,Neilsen12, Mineo2017}, suggesting that the QPO is correlated with the disc. However, this does not mean that the QPO definitely originates from the disc.

The results of the amplitude-ratio spectra suggest that the QPO originates from the corona. As the energy increases, some amplitude-ratios decrease and then flatten out for the two $\rho$ classes (Fig. \ref{fig:rmsratio}). This behaviour cannot be explained by assuming that the QPO originates from the disc. Considering that the disc emission is dominant at energy below $\sim 10$ keV while the corona emission is dominant at energy above $\sim 10$ keV \citep[e.g.][]{Gierlinski1999,Done07}, if the QPO is from the disc the amplitude-ratio cannot remain a constant but will ultimately increase as the energy increases above 10 keV due to the increasing ratio of the corona to the disc emission.

All of the amplitude-ratio spectra can be naturally explained by relating the QPO to the corona. If the QPO is from the corona, the amplitude-ratio will increase as the energy decreases below 10 keV, due to the increasing ratio of the disc to the corona emission, and it can remain more or less constant as the energy increases above 10 keV due to the difference in the amplitude spectrum of the QPO and the corona aperiodic variability.

The idea that the QPO originates from the corona is also consistent with the fact that the QPO absolute amplitude is low at the pulse phase (Fig. \ref{fig:lan.QPO}) where the corona flux is at its minimum \citep{Mineo2012, Mineo2017}, and with the result that both the QPO absolute amplitude and the corona flux at phase below $\sim 0.85$ are coincidentally higher in the $\rho_2$ class than in the $\rho_1$ class \citep[Fig. \ref{fig:lan.QPO}; Fig. 6 in][]{Neilsen12}. However, at phase above $\sim 0.85$ the QPO absolute amplitude is lower while the corona flux is still higher in the $\rho_2$ class, suggesting that the QPO absolute amplitude may depend upon other factors, rather than just the corona flux.

The correlation between the QPO and the disc can be qualitatively explained within the Lense-Thirring model of \citet{Ingram09}, who propose that the low-frequency QPOs of BHBs are the results of the Lense-Thirring precession of the hot inner flow (corona), and the QPO frequency is anti-correlated with the characteristic radius of the inner flow. In the two $\rho$ classes, the QPO frequency does not scale with the inner disc radius \citep[Fig. \ref{fig:lan.QPO}; Fig. 6 in][]{Neilsen12}, suggesting that the characteristic radius of the inner flow is different from the inner radius of the disc. However, in the scenario of the Lense-Thirring model, as the disc moves inwards, the QPO frequency increases due to the decrease of the characteristic radius of the inner flow, and meanwhile the disc temperature and the disc flux also increase. This may account for the correlation between the QPO frequency and the disc temperature and flux.

We discuss the physical implication of the QPO frequency spectra of the two $\rho$ classes on the basis of the Lense-Thirring model. In the context of the Lense-Thirring model, a positive correlation between the QPO frequency and energy implies that the corona photons with higher energy come from smaller radii, a non-correlation implies that the corona photons with different energies are from similar radii, and a negative correlation implies that the corona photons with lower energy are from smaller radii. The positive correlation in the slow rising phase of the two $\rho$ classes is naturally expected as the temperature of the accretion flow usually decreases with the radius \citep[e.g.][]{Shakura73, Mitsuda84, Zimmerman05, Vierdayanti13}. The non-correlation in the pulse and post-pulse phases may be due to the small size of the corona resulting from the cooling of the corona by the inward collapsing cold disc \citep{Neilsen11,Neilsen12}. The negative correlation of the SHQPO in the pulse phase of the $\rho_2$ class may be due to the larger size of the corona implied by the lower frequency of the SHQPO, and due to a faster cooling, especially for the innermost part of the corona, due to the higher accretion rate in the $\rho_2$ class than in the $\rho_1$ class \citep{Neilsen11,Neilsen12}. Note that the corona plasma which produces the SHQPO may be different from that which produces the QPO.

All of the QPO frequency spectra and the QPO phase-lags can be qualitatively explained using the model of \citet{Eijnden2016}, who suggest that the energy dependence of the QPO frequency, first found by \citet{Qu10}, is intrinsic to the QPO mechanism, and may result in the observed phase-lags. A positive correlation between QPO frequency and energy may result in a soft lag, while an anti-correlation between QPO frequency and energy may result in a hard lag. 

For the two $\rho$ classes, the QPO frequency spectra and the QPO phase-lags are presented in Fig. \ref{fig:40703fre} and in \citet{Yan2017}, respectively. For the QPO frequency spectra, the results of the least-squares fitting, the Monte Carlo simulation, and the Pearson correlation coefficient are consistent with each other, suggesting that the obtained slopes of the linear functions are credible.

In phases IV, V of the $\rho_1$ class and iii, iv, v of the $\rho_2$ class, the QPO frequency is positively correlated with energy, and the QPO phase-lag is soft. In phase III of the $\rho_1$ class, although the obtained slope is close to zero, the QPO frequency is significantly higher in the 20--40 keV than in the 2--40 keV band, and the QPO phase-lag is also soft. In phases II of the $\rho_1$ class and ii of the $\rho_2$ class, the QPO frequency is weakly correlated with energy and the QPO phase-lag is close to zero. In the phase vi of the $\rho_2$ class, the SHQPO frequency is anti-correlated with energy, and the phase-lag of the SHQPO is hard. All these results are consistent with the model of \citet{Eijnden2016}.

\subsection{disc--corona interaction}\label{sec:discuss2}

The high-frequency disc aperiodic variability may result from the disc--corona interaction. The high-frequency aperiodic variability produced by the disc may be smoothed out due to the dissipation or filter effect of the disc itself \citep{Revnivtsev99, Nowak99, Done07, Titarchuk07, Gierlinski08, Wilkinson09, Ingram11, Heil11}. The corona essentially has variability with broad-band frequency \citep[e.g.][]{Yan2017}. Then the high-frequency disc aperiodic variability detected through the amplitude-ratio spectrum method may be due to the disc--corona interaction, e.g. the corona affects the disc through wind/outflow and radiation.

The transition energy, $E_{\rm tr}$, of the amplitude-ratio spectrum can be used to investigate the formation region of the disc aperiodic variability and the disc--corona interaction. If the values of $E_{\rm tr}$ of the three amplitude-ratio spectra are similar, then it is expected that the disc aperiodic variabilities in these three frequency bands originate from similar disc radii, and that the whole X-ray radiation region of the disc may be influenced by the corona since the high-frequency disc aperiodic variability may result from the disc--corona interaction. If $E_{\rm tr}$ of the high-frequency amplitude-ratio spectrum is lower than that of the low- and intermediate-frequency ones, then the high-frequency disc aperiodic variability originates from larger radii, and at least part of the low- and intermediate-frequency disc aperiodic variability originates from smaller radii. This would also suggest that we have not detected the effect of the corona upon the innermost part of the disc.

For the pulse phase (VI) of the $\rho_1$ class, the 10--30 Hz amplitude-ratio spectrum is fitted with a linear function, indicating that the contribution of the disc is not significant for the 10--30 Hz aperiodic X-ray variability, and the corresponding $E_{\rm tr}$ should be very small. $E_{\rm tr}$ of the 10--30 Hz amplitude-ratio spectrum is thus lower than that of the 0.5--1 Hz and 1--10 Hz amplitude-ratio spectra for phase VI.

For the two $\rho$ classes, $E_{\rm tr}$ of the high-frequency amplitude-ratio spectrum is lower than that of the low- and intermediate-frequency ones in the pulse and post-pulse phase, and the values of $E_{\rm tr}$ of the three amplitude-ratio spectra are similar in the slow rising phase, except phase iv of the $\rho_2$ class. This suggests that the disc--corona interaction is different between the slow rising phase and the pulse/post-pulse phases. The value of $E_{\rm tr}$ of the high-frequency amplitude-ratio spectrum is lower than that of the low- and intermediate-frequency ones for phase iv of the $\rho_2$ class, but similar to them for phase IV of the $\rho_2$ class, suggesting that the disc--corona interaction is different between the two $\rho$ classes for part of the slow rising phase, although the two $\rho$ classes have a similar accretion process \citep{Neilsen12}.

The energy--frequency--phase-lag maps of the $\rho_1$ class present some interesting details of the disc--corona interaction (Fig. \ref{fig:FEL40703}). In all phase intervals except phase II, the phase-lag is obviously hard in the low-frequency band, implying that the disc, which is dominated by the low-frequency variability \citep{Yan2017}, drives the corona in all the phase intervals, except phase II during which some material is ejected from the inner part of the disc \citep{Neilsen11,Neilsen12}. In phase VI, the frequency range of the hard lag extends to above $\sim 2$ Hz, suggesting that when the disc collapses inwards \citep{Neilsen11,Neilsen12}, the disc drives the corona over a broader frequency band. The phase-lag is soft in the intermediate-frequency bands except the lower-left corner, indicating that the frequency range of the induced disc variability is narrower than that of the inducing corona variability, namely, there is ``frequency loss'' when the corona drives the disc. The soft phase-lag in the intermediate-frequency band is significant, especially in the slow rising phase, and the hard phase-lag in the low-frequency band is also significant, especially in the pulse phase.

We can also study the corona through the properties of the QPO, since the QPO originates from the corona. The coronae in the two classes may have similar properties but with different sizes or locations, which is suggested by the result that the count rate--QPO frequency relation and the count rate--QPO amplitude relation of the $\rho_1$ class are similar to those of the $\rho_2$ class, but with obvious shifts (Fig. \ref{fig:lanrate2fre2amp}). For the slow rising phase, the QPO frequency is higher in the $\rho_2$ class than in the $\rho_1$ class, suggesting that the characteristic size of the corona is smaller in the $\rho_2$ class based on the Lense-Thirring model. This is also implied by the fact that the hard lag during the slow rising phase is smaller in the $\rho_2$ class than in the $\rho_1$ class (Fig. \ref{fig:FEL40703}). The result that the size of the corona during the slow rising phase is smaller in the $\rho_2$ class than in the $\rho_1$ class, along with the result that both the QPO absolute amplitude and the corona flux are higher at the same time (see Section \ref{sec:discuss1}), providing us with interesting information for further discussing the property of the corona and the disc--corona interaction.

\section{CONCLUSIONS}\label{sec:con}

We present the results of the phase-resolved timing analysis of two {\it RXTE} observations of GRS 1915+105 during the single-peaked heartbeat ($\rho_1$ class) and the double-peaked heartbeat state ($\rho_2$ class), respectively. 

Although the QPO is correlated with the disc, as shown by the fact that the QPO frequency is correlated with the disc temperature and with the total flux that is dominated by the disc, the amplitude-ratio spectra strongly suggest that the QPO originates from the corona. The QPO frequency spectra can be qualitatively explained with the Lense-Thirring model of \citet{Ingram09}. The combination of QPO frequency spectrum and QPO phase-lag can be qualitatively interpreted using the model of \citet{Eijnden2016}.

The high-frequency disc aperiodic variability detected through the amplitude-ratio spectrum method may result from the disc--corona interaction. In some phases, the transition energies of the low-, intermediate-, and high-frequency amplitude-ratio spectra are similar, indicating that the disc aperiodic variability in these three frequency bands originates at similar disc radii, and the whole X-ray radiation region of the disc may be influenced by the corona. In some other phases, the transition energy of the high-frequency amplitude-ratio spectrum is lower than those of the low-frequency amplitude-ratio spectra, implying that we have not detected the effect of the corona on the innermost part of the disc.

The energy--frequency--phase-lag maps show that the disc, which is dominated by the low-frequency variability, always drives the corona, and the frequency range of the induced disc variability is narrower than that of the inducing corona variability when the corona drives the disc.

The differences in the count rate--QPO frequency relation, the count rate--QPO amplitude relation, and the phase-lag value imply that the corona in the two classes may have similar properties but with different sizes or locations.

\section*{Acknowledgements}

We thank the anonymous reviewer for comments which greatly improved the quality of the paper. We thank Saeqa Dil Vrtilek for helpful discussion. The research has made use of data obtained from the High Energy Astrophysics Science Archive Research Center (HEASARC), provided by NASA's Goddard Space Flight Center. This work is supported by National Natural Science Foundation of China (grant nos. 11273062, 11133001, 11333004, 11233001, 11233008, 11573071, and 11673023), the National Program on Key Research and Development Project (grant no. 2016YFA0400803), and China Postdoctoral Science Foundation (grant no. 2015M571838). LJ is also supported by the 100 Talents programme of Chinese Academy of Sciences.

\bibliographystyle{mn2e}
\bibliography{yan}

\label{lastpage}
\end{document}